\documentclass[10pt,a4paper,english]{amsart}
\usepackage[T1]{fontenc}
\usepackage[latin9]{inputenc}
\usepackage{color}
\usepackage{babel}
\usepackage{prettyref}
\usepackage{units}
\usepackage{amsthm}
\usepackage{amsmath}
\usepackage{amsbsy}
\usepackage{amstext}
\usepackage{amssymb}
\usepackage[unicode=true,pdfusetitle,
 bookmarks=true,bookmarksnumbered=false,bookmarksopen=false,
 breaklinks=false,pdfborder={0 0 0},backref=false,colorlinks=true]
 {hyperref}

\makeatletter

\pdfpageheight\paperheight
\pdfpagewidth\paperwidth

  \theoremstyle{plain}
  
  \theoremstyle{plain}
  \newtheorem*{prop*}{\protect\propositionname}
\theoremstyle{plain}

  \theoremstyle{plain}
  \newtheorem*{cor*}{\protect\corollaryname}

\usepackage{hyperref}
\usepackage{mathrsfs}
\usepackage{mathtools}
\usepackage{euscript}
\usepackage{upgreek}
\usepackage{algpseudocode}
\usepackage{algorithm}

\makeatother

  \providecommand{\corollaryname}{Corollary}
  \providecommand{\lemmaname}{Lemma}
  \providecommand{\propositionname}{Proposition}
\providecommand{\theoremname}{Theorem}

\theoremstyle{definition}

\begin{document}

\global\long\def\bbH{\mathbf{H}}
\global\long\def\fullv{X^{i_1}_{n_1} X^{i_2}_{n_2}\ldots X^{i_r}_{n_r}|v\rangle}
\global\long\def\rightv{X^{i_2}_{n_2}\ldots X^{i_r}_{n_r}|v\rangle}

\global\long\def\W{\mathcal{W}}
\global\long\def\lc{\mathopen{:}}
\global\long\def\rc{\mathclose{:}}
\global\long\def\bbC{\mathbf{C}}
\global\long\def\bbZ{\mathbf{Z}}
\global\long\def\bbR{\mathbf{R}}
\global\long\def\SL{\mathrm{SL}_{2}\!\left(\mathbb{Z}\right)}
\global\long\def\calM#1#2{\mathcal{M}_{#1}\!\left(#2\right)}
\global\long\def\sfM#1#2{\mathsf{M}_{#1}\!\left(#2\right)}
\global\long\def\sfS#1#2{\mathsf{S}_{#1}\!\left(#2\right)}
\global\long\def\calJ#1#2#3{\mathcal{J}_{#1}\!\left(#2,#3\right)}
\global\long\def\Ksum{\lim_{K\to\infty}\sum_{\gamma\in\Gamma_\infty\backslash\Gamma_{K,K^2}^*/\Gamma_\infty}}

\global\long\def\sm#1#2#3#4{\left(\begin{smallmatrix}#1  &  #2\cr\cr#3  &  #4\end{smallmatrix}\right)}
\global\long\def\dim{\mathrm{dim}}

\newcommand{\algoname}[1]{\textnormal{\textsc{#1}}}

\title[An Algorithm For Evaluating Gram Matrices]{An Algorithm for Evaluating Gram Matrices in Verma Modules of $\W$-algebras}

%
%
%
%


\maketitle

\begin{center}
         {Daniel Whalen}

\vspace{0.5cm} {\it SITP, Department of Physics and
Theory Group, SLAC,\\ Stanford University, Stanford, CA 94305, USA}\\
\end{center}

\begin{abstract}
I present a simple dynamic programming algorithm for the evaluation of operators in a wide range of superconformal algebras.  Special care is taken to describe the computation of the Gram matrix.  A \emph{Mathematica} package, \emph{Weaver.m}, is provided that implements the algorithm.\end{abstract}

\section{Introduction}

It is a generally unresolved question of how to determine the structure of highest-weight modules of $\mathcal W$-algebras.  I propose an algorithm that will efficiently perform evaluations on a Verma module at low level, enabling a number of explicit calculations.  These calculations include evaluating characters of irreducible representations, evaluating Gram matrices, constructing ladder diagrams, and finding primitive and p-singular vectors.

Related algorithms have appeared in previous work, but only for the Virasoro algebra~\cite{Feverati:2002vx}.  The algorithms described in this paper are applicable to general $\W$-algebras.

I provide an implementation of algorithms \ref{alg:Operator}, \ref{alg:algorithm2}, and \ref{alg:gram} as a \emph{Mathematica} package in the attached files.  Examples are provided for use of the package with the Virasoro algebra and the $\mathcal{SW}(3/2,2)$ algebra \cite{Gepner:2001cb,Shatashvili:1994uv}.

\section{Definitions}

A $\mathcal{W}$-algebra is an operator algebra that is generated by a finite number of fields and superfields, with explicit commutation relations and anti-commutation relations that are linear combination of single fields, normal ordered products of derivatives of two fields, and the operator describing the central extension of a Virasoro subalgebra\footnote{See, for example, \cite{Bouwknegt:1996tv} for a detailed discussion of $\W$-algebras.}.  We will construct an action of the $\W$-algebra on Verma modules of a finite-dimensional highest-weight space.

For a given $\W$-algebra, write the simple fields $X^i$, for $i=1,\ldots,d$  along with their conformal weights $\Delta_i$.  Write the parity $F^i$ of the field $X^i$, which is 0 if $X^i$ is bosonic, and is 1 if $X^i$ is fermonic.  If $X^i$ is bosonic, we write the corresponding set of operators $X^i_n$ for $n\in\bbZ$.  If $X^i$ is fermionic, we choose for each $i$, $n\in\bbZ$ (R~sector) or $n\in\bbZ+\frac{1}{2}$ (NS~sector).

The commutation and anti commutation relations can be written explicitly.  Let $[X^i_n,X^j_m]=X^i_nX^j_m-(-1)^{F^iF^j}X^j_mX^i_n$ and express the relations as 
\begin{equation*}
[X^i_n,X^j_m]=e^{ij}(n)\delta_{n+m}+\sum_{k} f^{ij}_k(n,m) X^k_{n+m} + \sum_{b,c,x,y} g^{ij}_{bc;xy}(n,m)\lc \partial^x X^b\partial^y X^c\rc_{n+m}.
\end{equation*}
The normal-ordered product is given in terms of $\partial^x X^i_n = x!\binom{-n-\Delta^i}x X^i_n$ by
\begin{equation*}
\lc \partial^xX^i\partial^yX^j\rc_n= \sum_{k\leq -\Delta_i-x}\partial^xX^i_k\partial^yX^j_{n-k}+(-1)^{F^iF^j}\sum_{k>-\Delta_i-x}\partial^yX^j_{n-k}\partial^xX^i_{n}.
\end{equation*}

The presence of normal-ordered products can be problematic for computational purposes.  Two additional assumptions can be made  regarding which $g$-functions are nontrivial in order for the relations to resolve nicely.  Construct the directed graph with a node for every unordered pair of operators $\{X^i,X^j\}$ and an edge from  $\{X^i,X^j\}$ to  $\{X^b,X^c\}$ if $g^{ij}_{bc;xy}(n,m)$ is not uniformly zero for all $n,m,x,$ and $y$.  We first require that this graph be acyclic.  We second require that if there exists a path from $\{X^i,X^j\}$ to $\{X^a,X^b\}$, we have $\max(i,j)>\min(a,b)$.  These requirements are sufficient, but not necessary conditions for the algorithm to work. The requirements will, however, apply to most algebras of interest up to a relabeling of the indices.

We are interested in constructing highest-weight representations of the $\W$-algebra.  Start with vectors $|v\rangle$ with $X^i_n|v\rangle=0$ whenever $n>0$.  The Verma module $M$ generated by the $|v\rangle$s is the span of vectors of the form $\fullv$.  Define a grading on the space such that one of the basis elements above has level $l=-\sum_{k=1}^r n_k$ and write the subspace of $M$ at level $l$ as $M_l$.  Any operator $X^i_n$ acts as a linear operator on each level: $X^i_n|_{M_p} =  X^i_n|_p:M_p\to M_{p-n}$.

$M_p$ has a basis that consists of states of the form $\fullv$ with the operators in the following order: the $n_i$s are in ascending order, with $i_{j+1}\geq i_j$ if $n_{i+1}=n_i$, and such that two identical fermionic operators never appear.  We call such a state properly ordered.

The $\W$-algebra has a 0 subalgebra that consists of operators of the form $X^i_0$.  Since ${X^i_0|_0:M_0\to M_0}$, $M_0$ forms a representation of the $0$-algebra.  Denote this algebra by $\rho$ and write $X^i_0|v\rangle = |\rho(X^i)\cdot v\rangle$.

%

We can construct a Hermitian conjugate on M.  On the highest weight space, we write $|v\rangle^\dagger = \langle v|$.  In general, we define ${X^i_n}^\dagger=X^i_{-n}$, although this may vary based on the construction of the $\W$-algebra.  Unitarity permits us to define an inner product on $M$.  Construct an inner product on $M_0$ that is unitary under the action of the 0-algebra.   The inner product can be extended to higher levels through the definition ${\langle\vec v,\vec w\rangle=\vec v^\dagger\vec w}$.

Observe that two vectors at different levels are necessarily orthogonal.  Let $(\vec v_1,\ldots,\vec v_r)$ be a basis for $M_l$.  We can therefore describe the inner-product completely by constructing the Gram matrix at level $l$, $G(l)$ where $G(l)_{ij} = \langle\vec v_i,\vec v_j\rangle$.

\section{Matrix Constructions}

For computational purposes, we will need to construct matrix representations of the above operators.  For a given level $l$, enumerate a properly ordered basis of $M_l$ as described above.  Write the dimensions as $\dim M_l=P(l)$.  Given a properly ordered state, $\vec w =\fullv$ at level $l$, write the $P(l)$-dimensional unit vector corresponding to $\vec w$ as $\overline{w}=\overline{\fullv}$.

The operators $X^i_n|_l$ can each be expressed as a $P(l-n)\times P(l)$ matrix acting on the left of $M_l$.  Denote this matrix by $\overline{X^i_n|_l}$.  Given a normal ordered product, $\lc X^i_nX^j_m\rc$, we express its action on $M_l$ as $\overline{\lc X^i_nX^j_m\rc |_l}$.  Note that when applied to a state at finite level, only a finite number of terms in the infinite sum defining the normal-ordered product survive, so this matrix is well-defined.  We also use the following shorthand:
\begin{equation*}
\overline{[X^i_n,X^j_m]|_l}=e(n)\delta_{n+m}I+\sum_{k} f^{ij}_k(n,m) \overline{X^k_{n+m}|_l} + \sum_{b,c,x,y} g^{ij}_{bc;xy}(n,m)\overline{\lc\partial^xX^b\partial^yX^c\rc_{n+m}|_l},
\end{equation*}
where it should be assumed that the expression on the left is replaced by the expression on the right whenever it appears in this paper.  Similarly, define as shorthand,
\begin{equation*}
\overline{((X^i_n X^j_m))|_l}=\begin{cases} 
\overline{X^i_n|_{l-m}}\cdot\overline{X^j_m|_l} &\mbox{if } $n<m$ \mbox{ or } $n=m$ \mbox{ and } i<j\\ 
\overline{[X^i_nX^j_m]|_l}+ (-1)^{F_iF_j}\overline{X^j_m|_{l-n}}\cdot\overline{X^i_n|_n} & \mbox{otherwise}. \end{cases} 
\end{equation*}
The expressions $\overline{((X^i_n X^j_m))|_l}$ and $\overline{X^i_n|_{l-m}}\cdot\overline{X^j_m|_l}$ are the same as operators, but may differ in the the order of evaluation.

Given a vector $\vec v\in M_l$, we can express $\vec v^\dagger$ as a $P(l)$-dimensional row-vector $\overline{v}^\dagger$ such that for $\vec w\in M_l$, $\vec v^\dagger\vec w = \overline{v}^\dagger\cdot\overline{w}$.  The unitary constraint then states that that
\begin{equation*}
\left(\overline{\fullv}\right)^\dagger= \left(\overline{\rightv}\right)^\dagger\cdot \overline{X^{i_1}_{-n_1}|_l}
\end{equation*}

The Gram matrix has a simple expression in this formalism.  The row corresponding to $\vec v$ is given in entirety by $\overline{v}^\dagger$.

Algorithms for evaluating the $\overline{X^i_n|_l},$ the $\overline{\lc X^i_nX^j_m\rc |_l}$ and the Gram matrices are given in Algorithms \ref{alg:Operator}, \ref{alg:algorithm2}, and \ref{alg:gram} respectively.

\begin{algorithm}
\caption{Evaluation of $\overline{X^i_n|_l}$}
\label{alg:Operator}
\begin{algorithmic}
\Function{Operator}{$i,n,l$}$=\overline{X^i_n|_l}$
	\If{ \Call{Operator}{$i,n,l$} has previously been evaluated}
		\State\Return\Call{Operator}{$i,n,l$}
	\EndIf
	\State $N\gets \{\}$
	\For{$\fullv$ in the basis of $M_l$}
		\State $row\gets$ \Call{OpAtState}{$i,n,l,\fullv$}
		\State $N\gets$ \Call{Append}{$N,row$}
	\EndFor
	\State\Return $N$
\EndFunction
\Statex
\Function{OpAtState}{$i,n,l,\fullv$}$=\overline{X^i_n|_l}\cdot\overline{\fullv}$
	\If{$l=0$ and $n=0$}
		\State\Return $\overline{|\rho(X^i_0)\cdot v\rangle}$
	\EndIf\If
	{$X^i_n\fullv$ is properly ordered}
		\State\Return $\overline{X^i_n\fullv}$
	\EndIf
	\If{$i=i_1$ and $n=n_1$ and $X^i$ is fermionic}
		\State\Return $\frac{1}{2}\overline{[X^i_n,X^i_n]|_{l+n}}\cdot\overline{\rightv}$
	\EndIf
	\If {$n>l+n_1$}
		\State $a\gets\overline{[X^i_n,X^{i_1}_{n_1}]|_{l+n_1}}\cdot\overline{\rightv}$
		\State\Return $a$
	\Else
		\State $a\gets \overline{[X^i_n,X^{i_1}_{n_1}]|_{l+n_1}}\cdot\overline{\rightv}$
		\State $b\gets (-1)^{F_iF_{i_1}}\overline{X^{i_1}_{n_1}|_{l+n_1-n}}\cdot\overline{X^{i}_{n}|_{l+n_1}}\cdot\overline{\rightv}$
		\State\Return $a+b$
	\EndIf	

\EndFunction
\end{algorithmic}
\end{algorithm}

\begin{algorithm}
\caption{Evaluation of $\overline{\lc \partial^xX^i\partial^yX^j\rc_n|_l}$}
\label{alg:algorithm2}
\begin{algorithmic}
\Function{NOP}{$i,j,x,y,n,l$}$=\overline{\lc \partial^xX^i\partial^yX^j\rc_n|_l}$
	\State $a\gets \sum\limits_{n-l\leq p\leq -\Delta_i}x!y!\binom{-p-\Delta_i}x\binom{-n+p-\Delta_j}y\overline{((X^i_p X^j_{n-p}))|_l}$
	\State $b\gets \sum\limits_{-\Delta_i<p\leq l}x!y!\binom{-p-\Delta_i}x\binom{-n+p-\Delta_j}y\overline{((X^j_{n-p} X^j_p))|_l}$
	\State\Return $a+b$
\EndFunction	
\Statex
\end{algorithmic}
\label{alg:NOperator}

\end{algorithm}

\begin{algorithm}
\caption{Evaluation of Gram Matrices}
\label{alg:gram}
\begin{algorithmic}
\Function{GramMatrix}{l}
	\If{l=0}
		\For{$i,j\in\{1,\ldots,P(0)\}$}
			\State $G_{ij}\gets \langle v_i|v_j\rangle$
		\EndFor
		\State\Return $G$
	\EndIf

	\For{$\fullv$ in the basis of $M_l$}
		\State $Row\gets\overline{\rightv}\cdot G(l+n_1)\cdot \overline{X_{-n_1}^{i_1}|_l}$
		\State $G\gets$ \Call{Append}{$N,Row$}
	\EndFor
	\State\Return G
\EndFunction	

\end{algorithmic}
\label{alg:Gram}
\end{algorithm}

\section{Consistency of the Algorithm}

The approach of the algorithm is straightforward.  \algoname{Operator} computes $\overline{X^i_n|_l}$ by using \algoname{OpAtState} to act on a basis for $M_l$.  If $X^i_n\fullv$ is properly ordered, then we can look up its value.  Otherwise, we need to commute the $X^i_j$ through $X^{i_1}_{n_1}$ and compute in terms of the other operators recursively.  

We demonstrate through induction over $l$ and $i$ that the recursion terminates.  For the moment, assume that normal ordered operators do not appear: that all the $g^{ij}_{bc;xy}$s are zero.  I delay extending the proof to normal ordered operators until the next section.

For the base case at $l=0$, \Call{OpAtState}{$i,n,0,|v\rangle$} returns an explicit expression line by line for each generator $|v\rangle$ of $M_0$. In particular,
\begin{equation*}
\Call{OpAtState}{i,n,0,|v\rangle} =\begin{cases}
\overline{|\rho(X^i_0)\cdot v\rangle} &\mbox{if } n = 0 \\
\overline{X^i_j|v\rangle} &\mbox{if } n < 0\end{cases}
\end{equation*}

Now proceed with the induction.  Assume inductively that $\overline{X^j_n|_p}$ has been evaluated for all $p<l$ and for all $p=l, j<i$.  We wish to show that $\overline{X^i_n|_l}$ can then be evaluated.  Consider the evaluation of \Call{OpAtState}{$i,n,l,\fullv$}.  There are only two cases that occur.  If  $X^i_n\fullv$ is properly ordered, the function immediately evaluates.  In the other case, we have $n_1<n$ or $n_1=n$ and $i_1<i$ or if $X^i$ is fermionic, we may have $n_1=n$ and $i_1=i$.   In these cases, \algoname{OpAtState} calls three operators, $\overline{[X^i_n,X^{i_1}_{n_1}]_{l+n_1}},\overline{X^{i_1}_{n_1}|_{l+n_1-n}},$ and $\overline{X^{i}_{n}|_{l+n_1}}$, all of which have already been evaluated by the assumption.

\section{Consistency of the Algorithm with Normal Ordered Operators}

Recall that a normal ordered product is defined in terms of the infinite sum
\begin{equation*}\label{nop}
\lc \partial^xX^i\partial^yX^j\rc_n= \sum_{k\leq -\Delta_i-x}\partial^xX^i_k\partial^yX^j_{n-k}+(-1)^{F^iF^j}\sum_{k>-\Delta_i-x}\partial^yX^j_{n-k}\partial^xX^i_{n},
\end{equation*}
of which only a finite number of terms survive when acting on a given level.  This expression is evaluated in \algoname{NOP} as

\begin{align*}\label{nopsum}
\overline{\lc \partial^x X^i\partial^yX^j\rc_n|_l}&=
 \sum\limits_{n-l\leq p\leq -\Delta_i}x!y!\binom{-p-\Delta_i}x\binom{-n+p-\Delta_j}y\overline{((X^i_p X^j_{n-p}))|_l}\\
 &\phantom{aaaaaa}+\sum\limits_{-\Delta_i<p\leq l}x!y!\binom{-p-\Delta_i}x\binom{-n+p-\Delta_j}y\overline{((X^j_{n-p} X^j_p))|_l}
\end{align*}

Return to the claim in the last chapter that the recursion in \algoname{Operator} terminates.  I wish to extend the result to instances where the normal ordered products appear.  Assume, as before, that we have calculated all the operators $\overline{X^j_n|_p}$ for $p<l$ and for ${p=l},{j<i}$.  We wish to then evaluate $\algoname{OpAtState}$ for $\overline{X^i_n|_l}$ acting on $\overline{\fullv}$.

Recall the assumptions that we made regarding the $g^{ij}_{bc;xy}(n,m)$.  We constructed the directed graph with an edge from  $\{X^i,X^j\}$ to  $\{X^b,X^c\}$ if $g^{ij}_{bc;xy}(n,m)$ is not uniformly zero for all $n,m,x,$ and $y$.  We assumed that this graph is acyclic and that paths may only exist from $\{X^i,X^j\}$ to $\{X^a,X^b\}$ if $\max(i,j)>\min(a,b)$.

$\algoname{OpAtState}$ only introduces a normal-ordered product if $X^i_n$ and $X^{i_1}_{n_1}$ need to be commuted, when $n>n_1$ or $n=n_1$ and $i>i_1$.  In this case, a normal ordered product will be introduced that depends on terms of the form $\overline{((X^j_{n} X^j_m))|_l}$.  We can expand out the commutator in that term which may give a normal-ordered product.  Iteratively repeat this process. The iteration terminates because of the acyclic condition we imposed on the normal-ordered dependency graph.

When we are done, we are left with terms that involve operators of the form $\overline{X^k_{n_1+n}|_{l+n_1}}$ and possibly of the form $\overline{X^a_{n+n_1-p}|_{l+n_1-p}}\cdot \overline{X^b_{p}|_{l+n_1}}$ if there is a path from $(n,n_1)$ to $(a,b)$ in the normal-ordered dependency graph and if $p>n+n_1-p$ or $p=n+n_1-p$ and $b>a$.  Operators of the first form have all been evaluated by the induction hypothesis.  Operators of the second form can only fail to have been evaluated if $p=n_1=n=n+n_1-p$, $a<b$ and if $a\geq n$.  Since $a=\min(a,b)$ and $n=\max(n,n_1)$, this is impossible by our second condition on the normal-ordered dependency graph.  Thus, the recursion in the algorithm terminates.

\section{Computational Complexity}

\subsection{Asymptotics of $P(l)$}
The number of states at a given level $l$ has an asymptotic expression, which is a slight generalization of the asymptotics of the partition function.  Let $N_B$ be the number of bosonic operators, $N_F$ be the number of fermionic operators, $\alpha = {\pi\sqrt{N_F+2N_B}}/{\sqrt{3}}$, and $\beta = -(N_B+3)/4$.  Then the number of states at level $l$ is \cite{Nicolas:2000bx,Nicolas:2003ih}
\begin{equation*}
P(l)=\dim M_l  = O\left(l^\beta e^{\alpha\sqrt{l}} \right).
\end{equation*}

We will make frequent use of the asymptotic form of sums over the partition functions, $\sum_{p=0}^l  p^\beta e^{\alpha\sqrt{p}} = \frac{2}{\alpha}l^{\beta+\frac{1}{2}} e^{\alpha\sqrt{l}}+O(l^\beta e^{\alpha\sqrt{l}})$.

\subsection{Matrix multiplication}
Multiplication of two $n\times n$ square matrices can be performed in $O(n^\theta)$ time for $2\leq\theta<3$.  The best known algorithms give $\theta\approx2.373$, although in practice, textbook matrix multiplication with $\theta=3$ or the Strassen algorithm with $\theta\approx2.81$ will be employed.

All of these algorithms can be applied to rectangular matrices by subdividing the matrices into blocks and performing square matrix multiplication on the subblocks.  Doing so allows for the multiplication of an $l\times m$ and an $m\times n$ matrix to be performed in $O(lmn\cdot\min(l,m,n)^{\theta-3})$ time.

\subsection{Operators}

Consider the problem of evaluating all of the operators $\overline{X^i_n|_p}$ such that $0\leq p\leq l$, and $-l\leq n\leq l$ subject to the constraint that  ${0\leq p-n\leq l}$.  For calculating a single operator $\overline{X^i_n|_p}$, assume that the matrix expressions for lower level operators and normal ordered operators have been evaluated.  The evaluation of the normal-ordered products will be treated in the next subsection.

The slowest step in \algoname{OpAtState} is the evaluation of $b$, since all of the other operators are multiplication of a matrix by a unit vector or the lookup of the unit vector corresponding to a state, both of which can be perfumed in slightly over $O(P(l))$ time.  Both cases have an upper bound of time $P(p)P(p-m)$, so the net complexity of the operation is $P(p)^2 P(p-n)$.  Summing over $p$ and $n$, the total time is $l P(l)^3$.

There is a modification to the algorithm that decreases the complexity.  Make the following modification to Algorithm \ref{alg:Operator}.  Instead of evaluating the $P(p)$ rows separately as in \algoname{Operator}, we evaluate all of the rows $\fullv$ that share the same $X^{i_1}_{n_1}$ together.  For a fixed $X^{i_1}_{n_1}$, write $N(X^{i_1}_{n_1})$ for the submatrix of $N$ from \algoname{Operator} whose rows correspond with the states starting with $X^{i_1}_{n_1}$.  When calculating $b$ in \algoname{OpAtState}, multiple vector multiplications may then be turned into a single matrix multiplication.  A similar calculation gives an upper bound of the runtime of this modified algorithm as $O(l^{\frac{3}{2}}P(l)^\theta)$.

\subsection{Normal Ordered Operators}
In order to evaluate the normal ordered operators $\overline{\lc X^iX^j\rc_n|_p}$, the normal ordered product needs to be expanded into a sum of products of operators from levels $p$ to $k$ and from $k$ to $p-n$ as $k$ varies from 0 to $p-\frac{n}{2}$.  The products can be calculated in $O(P(p)P(k)P(p-n) P(\min(p,k,p-n))^{\theta-3})$ time.  
Summing over $k, p$ and $n$, the total time is $O(l^\frac{3}{2} P(l)^\theta)$.

\subsection{Gram Matrices}
Having evaluated all of the operators and normal ordered products of operators, we can evaluate the Gram matrix.  The Gram matrix has $P(l)$ rows, each of which can be evaluated recursively from a previous Gram matrix and a operator of size less then $P(l)\times P(l)$.  Evaluating the Gram matrices for level 0 through $l$ can therefore be performed in $O(\sqrt{l}P(l)^3)$ once the operators have been computed.

The vector multiplications can be merged in the evaluation of $G(l)$ in Algorithm~\ref{alg:gram} similarly to the evaluation of \algoname{Operator}.  Doing so will reduce the complexity of the Gram matrix evaluations to be less than that for evaluating \algoname{Operator}.

\subsection{Total Complexity}

The above calculations provide an upper bound for the total run time of $O(l^\frac{3}{2}P(l)^\theta)$ to evaluate all of the operators up to level $l$.

\bibliography{Algorithm_Details}{}
\bibliographystyle{plain}

\end{document}